# A systematic approach towards robust stability analysis of integral delay systems with general interval kernels

**H. Taghavian**

KTH Royal Institute of Technology, Stockholm, Sweden.
Email: hamedta@kth.se

*Abstract*— robust stability problem of integral delay systems with uncertain kernel matrix functions is addressed in this paper. On the basis of characteristic equation and the argument principle, an algorithm is generated which is shown to outperform the Lyapunov-Krasovskii (LK) approaches with respect to conservatism in the presented examples. Despite the conventional manual use of Nyquist criterion, the proposed algorithm is fully algebraic, cheaper and easily implemented in computer programs. In addition, the proposed method is applicable to a wider range of uncertain systems compared to the existing literature. Namely, despite the previously published results on this problem, the kernel matrix function here is not limited to exponential type and can include any real function within known bounds as its elements.

## 1. Introduction

Just like uncertainties in control systems (Taghavian & Tavazoei (2017), (2018) and Vafamand et al. (2019)), time delays are also omnipresent in various engineering systems, such as communication channels (Berry & Gallager (2002)) and chemical processes (Mehrkanoon et al. (2016)). Not surprisingly, significant attention has therefore been paid to study of time-delay systems suffering from uncertainties (Yan et al. (2019), Sun et al. (2019) and Si et al. (2019)). Integral delay dynamics naturally appear in the context of various common control problems in the presence of time-delays. For some of the important samples in this regard, see Melchor-Aguilar et al. (2010). It is well known that control systems suffering from time delays can be naturally infinite dimensional having transcendental characteristic equations that cannot be expressed via elementary functions. Things get even worse for the class of integral delay systems whose characteristic equations can virtually be any irrational function with a complex domain. This has considerably inhibited the usual frequency domain analysis methods for these systems. Lyapunov-Krasovskii (LK) approaches somewhat make up for this shortage by formulating stability conditions in some LMIs (Li et al. (2016), Melchor-Aguilar (2016) and Zhou & Li (2016)). However, they are notorious for increased conservatism. This hints on the importance of developing stability analysis methods based on the characteristic equations of these systems which although rarely, is considered in some papers in the literature (Li et al. (2016) and (2013)).

Based on the above argument, this paper aims at robust stability analysis of integral delay systems suffering from uncertainty in the kernel. This problem has been previously investigated in different frameworks using LK functionals (Melchor-Aguilar & Morales-Sánchez (2016), Melchor-Aguilar et al. (2008), Morales-Sánchez & Melchor-Aguilar (2013)). However in order to reduce the conservatism involved in the LK-based approaches, the aforementioned problem is met in the frequency domain with help of the argument principle in this paper. Particularly, inspired by the robust Nyquist arrays of multivariable systems (Chen & Seborg (2002)), an algorithm is constructed on the basis of characteristic equations of integral delay systems, by utilizing eigenvalue inclusion sets of partition matrices, eigenvalue bounding techniques of interval matrices and Fourier companion matrices. Since a usual evaluation of the Nyquist diagram of an integral delay system tends to be cumbersome, an algorithm is then designed to determine stability by exploiting the Nyquist diagram information only at some critical frequency points, obviating the need to plot anything. The critical idea in this method is detecting encirclements of the Nyquist plot in an algebraic root finding program and bounding the uncertain kernel entries by spline functions to produce a periodic behavior in the region containing the Nyquist diagram of the uncertain system and hence significantly reducing the computation burden eventually.

This paper is organized as follows: In the next section, the problem is propounded mathematically. Section 3 is devoted to robust stability analysis foundations in the frequency domain following the main results. All the main results are then utilized to provide an algorithm for a convenient use in section 4. Section 5 presents numerical evaluation of the results and finally, conclusion remarks are provided in section 6. In this paper, the following conventions hold: matrix elements are distinguished by indices $i'$ and $j'$, notation $\lambda_i\{.\}$ is used to refer to the $i^{th}$ eigenvalue of a matrix, $L_{t \to s}\{.\}$ ($L_{s \to t}^{-1}\{.\}$) is used for the (inverse) Laplace transform and $j = \sqrt{-1}$.

## 2. Preliminaries

Consider a general integral delay system defined as follows:

$$x(t) = \int_0^{\bar{\tau}} A(\tau) x(t-\tau) d\tau, \quad t > \bar{\tau} \tag{1}$$

in which $x: \mathbb{R}^{\geq 0} \to \mathbb{R}^n$ denotes the state vector with the initial function $\varphi: [0, \bar{\tau}] \to \mathbb{R}^n$ satisfying $x(t) = \varphi(t)$, $0 \leq t \leq \bar{\tau}$. In (1) let the kernel matrix $A: [0, \bar{\tau}] \to \mathbb{R}^{n \times n}$ be unknown lying between the two known matrices $\bar{A}(\tau) = [\bar{a}_{i',j'}(\tau)]_{n \times n}$ and $\underline{A}(\tau) = [\underline{a}_{i',j'}(\tau)]_{n \times n}$ as:

$$\underline{A}(\tau) \leq A(\tau) \leq \bar{A}(\tau) \tag{2}$$

Inequality (2) holds element-wise implying:

$$\underline{a}_{i',j'}(\tau) \leq a_{i',j'}(\tau) \leq \bar{a}_{i',j'}(\tau) \tag{3}$$

for $\forall \tau \in [0, \bar{\tau}]$ and $i', j' \in \{1, \cdots, n\}$, where $a_{i',j'}(\tau)$ denotes the entries of the kernel matrix $A(\tau)$. The bounding functions $\underline{a}_{i',j'}(\tau)$ and $\bar{a}_{i',j'}(\tau)$ are considered to be in the following spline form:

$$\begin{cases} \bar{a}_{i',j'}(\tau) = \sum_{k=0}^{N-1} \bar{b}_{i',j',k} p_{n_0,k}(\tau) \\ \underline{a}_{i',j'}(\tau) = \sum_{k=0}^{N-1} \underline{b}_{i',j',k} p_{n_0,k}(\tau) \end{cases} \tag{4}$$

in which $N = \bar{\tau}/h$ is a natural number and

$$p_{n_0,k}(\tau) = (\tau - kh)^{n_0} H(\tau - kh) - (\tau - (k+1)h)^{n_0} H(\tau - (k+1)h) \tag{5}$$

where $H(.)$ is the Heaviside step function. The purpose of this paper is to investigate robust stability problem of system (1) in the presence of uncertainty in the kernel matrix. To this aim, one needs to study the characteristic equation of (1) as:

$$\Delta(s) = \det(I_n - M(s)) = 0 \tag{6}$$

in which $M(s) = \int_0^{\bar{\tau}} A(\tau) \exp(-s\tau) d\tau$ is the Laplace transform of the kernel with entries $m_{i',j'}(s)$. System (1) is stable if and only if all the roots of (6) are located in the open left half plane. With help of the argument principle, it can be shown that this condition can also be interpreted as stated in the following lemma.

**Lemma 1 (Maciejowski (1989))**

Assume that matrix $I_n - M(j\omega)$ is nonsingular for $\omega \in \mathbb{R}$. The number of unstable characteristic roots of (1) equals the total number of clockwise encirclements of the point $+1$ on the complex plane by eigenvalues $\lambda_i\{M(j\omega)\}$ ($i = 1, 2, \cdots, n$) where $\omega \in \mathbb{R}$.

Foundations of robust stability analysis are constructed in the next section, followed by the main results of the paper.

## 3. Main results

Lemma 1 provides a graphical means of stability analysis which is similar to the Nyquist arrays technique in MIMO LTI systems analysis. Prior to doing so however, one needs to capture the possible location of eigenvalues of the uncertain system on the complex plane which is eventually realized in Lemma 3. The first step on this matter is to quantify the uncertainty as it undergoes the Fourier transform $M(j\omega) = L_{\tau \to s}\{A(\tau)\}_{|s=j\omega}$. This is the main concern of the following lemma.

**Lemma 2**

Let $a_{i',j'}(\tau)$ be an uncertain function satisfying (3) and define:

$$\begin{cases} \tilde{m}_{i',j'} = \frac{1}{2}\int_0^{\bar{\tau}}\left(\bar{a}_{i',j'}(\tau) - \underline{a}_{i',j'}(\tau)\right)d\tau \\ \hat{m}_{i',j'}(j\omega) = \frac{1}{2}\left(\overline{m}_{i',j'}(j\omega) + \underline{m}_{i',j'}(j\omega)\right) \end{cases} \quad (7)$$

where

$$\begin{cases} \overline{m}_{i',j'}(j\omega) = L_{\tau \to s}\{\bar{a}_{i',j'}(\tau)\}_{|s=j\omega} \\ \underline{m}_{i',j'}(j\omega) = L_{\tau \to s}\{\underline{a}_{i',j'}(\tau)\}_{|s=j\omega} \end{cases}$$

Then $m_{i',j'}(j\omega) = L_{\tau \to s}\{a_{i',j'}(\tau)\}_{|s=j\omega}$ is located within a square on the complex plane, centered at $\hat{m}_{i',j'}(j\omega)$ with the side lengths $2\tilde{m}_{i',j'}$ for $\forall \omega \in \mathbb{R}$.

Proof:

Inequality (3) can be converted into the following equality by using the unknown function $c_{i',j'}(\tau)$:

$$a_{i',j'}(\tau) + c_{i',j'}(\tau) = \bar{a}_{i',j'}(\tau) \quad (8)$$

which satisfies

$$0 \leq c_{i',j'}(\tau) \leq \bar{a}_{i',j'}(\tau) - \underline{a}_{i',j'}(\tau) \quad (9)$$

Based on (8), one easily has:

$$\text{Re}\{m_{i',j'}(j\omega)\} = \text{Re}\{\overline{m}_{i',j'}(j\omega)\} - \int_0^{\bar{\tau}} c_{i',j'}(\tau)\cos(\omega\tau)d\tau \quad (10)$$

for the real part of $m_{i',j'}(j\omega)$. According to (9) and noting the inequality

$$\frac{1}{2}(\cos(\omega\tau) + 1) \geq \max_{0\leq\tau\leq\bar{\tau}}\{0, \cos(\omega\tau)\} \quad (11)$$

it is deduced that:

$$\int_0^{\bar{\tau}} c_{i',j'}(\tau)\cos(\omega\tau)d\tau \leq$$
$$\frac{1}{2}\int_0^{\bar{\tau}}\left(\bar{a}_{i',j'}(\tau) - \underline{a}_{i',j'}(\tau)\right)(\cos(\omega\tau) + 1)d\tau \quad (12)$$
$$= \frac{1}{2}\left(\text{Re}\{\overline{m}_{i',j'}(j\omega)\} - \text{Re}\{\underline{m}_{i',j'}(j\omega)\}\right) + \tilde{m}_{i',j'}$$

where $\tilde{m}_{i,j}$ is given by (7). Likewise, by using (9) and the inequality

$$\frac{1}{2}(\cos(\omega\tau) - 1) \leq \min_{0\leq\tau\leq\bar{\tau}}\{0, \cos(\omega\tau)\} \quad (13)$$

it is obtained that:

$$\int_0^{\bar{\tau}} c_{i',j'}(\tau)\cos(\omega\tau)d\tau \geq$$
$$\frac{1}{2}\left(\text{Re}\{\overline{m}_{i',j'}(j\omega)\} - \text{Re}\{\underline{m}_{i',j'}(j\omega)\}\right) - \tilde{m}_{i',j'} \quad (14)$$

Using inequalities (12) and (14) in (10) yields in:

$$\left|\text{Re}\{m_{i',j'}(j\omega) - \hat{m}_{i',j'}(j\omega)\}\right| \leq \tilde{m}_{i',j'} \quad (15)$$

where $\hat{m}_{i',j'}(j\omega)$ is given by (7). Following a rather similar procedure, the following inequality can also be derived for the imaginary part of $m_{i',j'}(j\omega)$:

$$\left|\text{Im}\{m_{i',j'}(j\omega) - \hat{m}_{i',j'}(j\omega)\}\right| \leq \tilde{m}_{i',j'} \quad (16)$$

Inequalities (15) and (16) together prove this lemma.

∎

Based on Lemma 2, the uncertainty in the kernel matrix can be translated into the frequency domain by using the following set of element-wise matrix inequalities:

$$\begin{cases} -\tilde{M} \leq \text{Re}\{M(j\omega)\} - \text{Re}\{\hat{M}(j\omega)\} \leq \tilde{M} \\ -\tilde{M} \leq \text{Im}\{M(j\omega)\} - \text{Im}\{\hat{M}(j\omega)\} \leq \tilde{M} \end{cases} \quad (17)$$

where $\hat{M}(j\omega) = \left[\hat{m}_{i',j'}(j\omega)\right]_{n\times n}$, $\tilde{M} = \left[\tilde{m}_{i',j'}\right]_{n\times n}$ and the entries $\hat{m}_{i',j'}(j\omega)$, $\tilde{m}_{i',j'}$ are given in (7). Based on this result and Lemma 1, a sufficient condition of instability is presented in the following theorem by investigating $\hat{M}(j\omega)$ only at $\omega = 0$. This theorem indicates that system (1) can usually be proved unstable by quickly checking a simple condition, if there are an odd number of unstable characteristic roots associated with it. This is an independent side result and is not incorporated in the main algorithm presented in section 4.

**Theorem 1**

Define:

$$R = \hat{M}_J - diag\left(\lambda_1\{\hat{M}(0)\}, \cdots, \lambda_n\{\hat{M}(0)\}\right) + |V^{-1}||\tilde{M}||V|$$

where $\hat{M}_J = V^{-1}\hat{M}(0)V$ is the Jordan form of $\hat{M}(0)$ and $|.|$ denotes the corresponding matrix with modulus elements. Consider the circles with centers $\lambda_i\{\hat{M}(0)\}$ ($i=1,2,\cdots,n$) and radii $\hat{r}_i = \sum_{i'=1}^{n} r_{i,i'}$ (or $\hat{r}'_i = \sum_{i'=1}^{n} r_{i',i}$) where $r_{i',j'}$ denotes the elements of matrix $R$. System (1) is unstable, if an odd number of eigenvalues $\lambda_i\{\hat{M}(0)\}$ are situated on the ray $\{x+iy \mid x>1\}$ and the unions of the circles described above joint with the ray $\{x+iy \mid x>1\}$ are disjoint from the ray $\{x+iy \mid x<1\}$.

Proof:

Since according to Riemann–Lebesgue lemma, all eigenvalues of $M(j\omega)$ tend to origin as $\omega \to \infty$, it is deduced that all the eigenvalues of $M(j\omega)$ that are located on the ray $\{x+iy \mid x<1\}$ at $\omega=0$ would eventually make an even number of encirclements of the point $+1$ on the complex plane. Also as $M(0) \in \mathbb{R}^{n\times n}$, each pair of eigenvalues of $M(j\omega)$ which make complex conjugates at $\omega=0$ would also make an even number of encirclements together, as $\omega$ tends from $-\infty$ to $+\infty$. In contrast, each eigenvalue of $M(j\omega)$ that is on the ray $\{x+iy \mid x>1\}$ at $\omega=0$ makes an odd number of encirclements. Summing up, Considering Lemma 1, it is realized that system (1) has an odd number of unstable characteristic roots if an odd number of $\lambda_i\{M(0)\}$ are located on the ray $\{x+iy \mid x>1\}$. In order to follow this rationale however, since $M(0)$ is an interval matrix, its modified Gershgorin circles introduced in Juang & Shao (1989) (and in the statement of this theorem) are used here to determine the possible location of its eigenvalues.

∎

The next step before applying Lemma 1 would be specification of a region on the complex plane encompassing the unknown eigenvalues of $M(j\omega)$, which is worked out in next lemma.

**Lemma 3**

Consider the matrices $\hat{M}(j\omega)$ and $\tilde{M}$ whose elements are defined in (7) and define:

$$T = \begin{bmatrix} 1 & 1 \\ 1 & 1 \end{bmatrix} \otimes (\tilde{M} + \tilde{M}^T)$$

$$\delta_R(\omega) = \rho\{T\} + ((2n-1)/n)^{1/2} \times$$
$$\left(tr\left((\hat{M}_R(\omega)+\hat{M}_R^T(\omega))^2 - (\hat{M}_I(\omega)-\hat{M}_I^T(\omega))^2\right) - 4\left(tr(\hat{M}_R(\omega))\right)^2 / n\right)^{1/2}$$

$$\delta_I(\omega) = \rho\{T\} + ((2n-1)/n)^{1/2} \times$$
$$\left(tr\left((\hat{M}_I(\omega)+\hat{M}_I^T(\omega))^2 - (\hat{M}_R(\omega)-\hat{M}_R^T(\omega))^2\right) - 4\left(tr(\hat{M}_I(\omega))\right)^2 / n\right)^{1/2}$$

(18)

where $\otimes$ denotes the Kronecker product and $\hat{M}_R(\omega)$ and $\hat{M}_I(\omega)$ are shorthand notations for the real and imaginary parts of $\hat{M}(j\omega)$ respectively. All the eigenvalues $\lambda_i\{M(j\omega)\}$ ($i=1,2,\cdots,n$, $\omega \in \mathbb{R}$) are located within the band $Q = \bigcup_{\omega \in \mathbb{R}} Q(\omega)$ where $Q(\omega)$ is a rectangle centered at $Q_C(\omega) = tr(\hat{M}(j\omega))/n$ with the vertical and horizontal side lengths of $\delta_I(\omega)$ and $\delta_R(\omega)$ respectively.

Proof:

Observe that $M(j\omega) \in \mathbb{C}^{n\times n}$ is uncertain satisfying the inequalities set (17) element-wise. Since determination of the possible location of eigenvalues of a perturbed matrix has constantly been a hot topic in linear algebra due to its vast range of applications in robust control, many techniques have been developed so far to handle this kind of problem. Utilizing one of the most efficient and least conservative results in this regard offered by Hladík (2013) reveals that the real and imaginary parts of the eigenvalues of $M(j\omega)$ satisfy the following inequalities:

$$\min_i \lambda_i \{S_R(\omega)\} - \rho\{T\} \leq 2\operatorname{Re}\{\lambda_i\{M(j\omega)\}\} \leq$$
$$\max_i \lambda_i \{S_R(\omega)\} + \rho\{T\}$$
$$\min_i \lambda_i \{S_I(\omega)\} - \rho\{T\} \leq 2\operatorname{Im}\{\lambda_i\{M(j\omega)\}\} \leq \quad (19)$$
$$\max_i \lambda_i \{S_I(\omega)\} + \rho\{T\}$$

where $\rho\{.\}$ denotes the spectral radius and

$$S_R(\omega) = \begin{bmatrix} \hat{M}_R(\omega) + \hat{M}_R^T(\omega) & \hat{M}_I^T(\omega) - \hat{M}_I(\omega) \\ \hat{M}_I(\omega) - \hat{M}_I^T(\omega) & \hat{M}_R(\omega) + \hat{M}_R^T(\omega) \end{bmatrix}$$
$$S_I(\omega) = \begin{bmatrix} \hat{M}_I(\omega) + \hat{M}_I^T(\omega) & \hat{M}_R(\omega) - \hat{M}_R^T(\omega) \\ \hat{M}_R^T(\omega) - \hat{M}_R(\omega) & \hat{M}_I(\omega) + \hat{M}_I^T(\omega) \end{bmatrix} \quad (20)$$

With aim to derive simpler stability conditions at the expense of increased conservatism, one may approximate the eigenvalues of symmetric matrices $S_R(\omega)$ and $S_I(\omega)$. Using the inclusion region particularly designed for partition matrices in Zou & Jiang (2010) indicates that all the eigenvalues of matrices $S_R(\omega)$ and $S_I(\omega)$ are included in two closed disks centered at $2\operatorname{tr}(\hat{M}_R(\omega))/n$ and $2\operatorname{tr}(\hat{M}_I(\omega))/n$ with the radii:

$$\sqrt{\frac{2n-1}{2n}\left(\|S_R(\omega)\|_F^2 - 8\left(\operatorname{tr}(\hat{M}_R(\omega))\right)^2/n\right)} \quad (21)$$

and

$$\sqrt{\frac{2n-1}{2n}\left(\|S_I(\omega)\|_F^2 - 8\left(\operatorname{tr}(\hat{M}_I(\omega))\right)^2/n\right)} \quad (22)$$

respectively, where $\|.\|_F$ denotes the Frobenius matrix norm. This implies:

$$\left|\lambda_i\{S_R(\omega)\} - 2\operatorname{tr}(\hat{M}_R(\omega))/n\right| \leq$$
$$\sqrt{\frac{2n-1}{2n}\left(\|S_R(\omega)\|_F^2 - 8\left(\operatorname{tr}(\hat{M}_R(\omega))\right)^2/n\right)}$$
$$\left|\lambda_i\{S_I(\omega)\} - 2\operatorname{tr}(\hat{M}_I(\omega))/n\right| \leq \quad (23)$$
$$\sqrt{\frac{2n-1}{2n}\left(\|S_I(\omega)\|_F^2 - 8\left(\operatorname{tr}(\hat{M}_I(\omega))\right)^2/n\right)}$$

for $i \in \{1, \cdots, 2n\}$. Considering definitions (20) and rewriting the norms $\|S_R(\omega)\|_F^2$ and $\|S_I(\omega)\|_F^2$ in (23) in terms of matrix traces, one concludes

$$\left|\lambda_i\{S_R(\omega)\} - 2\operatorname{tr}(\hat{M}_R(\omega))/n\right| \leq ((2n-1)/n)^{1/2} \times$$
$$\left(\operatorname{tr}\left((\hat{M}_R(\omega) + \hat{M}_R^T(\omega))^2 - (\hat{M}_I(\omega) - \hat{M}_I^T(\omega))^2\right) - \right.$$
$$\left. 4\left(\operatorname{tr}(\hat{M}_R(\omega))\right)^2/n\right)^{1/2}$$
$$\left|\lambda_i\{S_I(\omega)\} - 2\operatorname{tr}(\hat{M}_I(\omega))/n\right| \leq ((2n-1)/n)^{1/2} \times \quad (24)$$
$$\left(\operatorname{tr}\left((\hat{M}_I(\omega) + \hat{M}_I^T(\omega))^2 - (\hat{M}_R(\omega) - \hat{M}_R^T(\omega))^2\right) - \right.$$
$$\left. 4\left(\operatorname{tr}(\hat{M}_I(\omega))\right)^2/n\right)^{1/2}$$

Using the lower and upper bounds of the eigenvalues of $S_R(\omega)$ and $S_I(\omega)$ given by inequalities (24) in inequalities (19) gives:

$$-\delta_R(\omega)/2 \leq \operatorname{Re}\{\lambda_i\{M(j\omega)\}\} -$$
$$\operatorname{Re}\{\operatorname{tr}(\hat{M}(j\omega))\}/n \leq \delta_R(\omega)/2$$
$$-\delta_I(\omega)/2 \leq \operatorname{Im}\{\lambda_i\{M(j\omega)\}\} - \quad (25)$$
$$\operatorname{Im}\{\operatorname{tr}(\hat{M}(j\omega))\}/n \leq \delta_I(\omega)/2$$

which proves this lemma. ∎

In order to visualize Lemma 3, the band $Q$ associated with system (1) with sample parameters $n = 2$, $n_0 = 1$, $h = 0.5$, $\bar{\tau} = 1$, $\bar{b}_{1,1,0} = 7.6$, $\bar{b}_{1,1,1} = -7.4$, $\bar{b}_{2,1,0} = 1.6$, $\bar{b}_{2,1,1} = -1.4$, $\bar{b}_{1,2,0} = 1.1$, $\bar{b}_{1,2,1} = -0.9$, $\bar{b}_{2,2,0} = 7.1$, $\bar{b}_{2,2,1} = -6.9$ and $\underline{b}_{i',j',k} = \bar{b}_{i',j',k} - 0.2$ is plotted in figure 1.

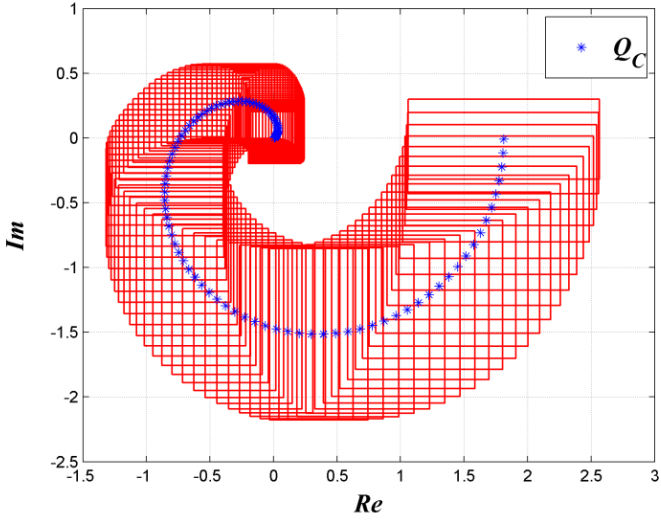

**Figure 1:** the band $Q$ associated with system (1) with some sample parameters.

Having the possible region of the system eigenvalues on the complex plane in Lemma 3, one may proceed to use Lemma 1 to check robust stability of system (1) with an uncertain kernel $A(\tau)$. Rather than directly doing so however, the whole procedure is studied and manipulated to make it more convenient and less costly by taking advantage of general system properties. In order to show this, firstly note that since the bounding functions $\overline{a}_{i',j'}(\tau)$ and $\underline{a}_{i',j'}(\tau)$ are in spline form, the elements of matrix $\hat{M}(j\omega)$ (defined in (7)) can be computed explicitly as:

$$\hat{m}_{i',j'}(j\omega) = \frac{n_0!}{2(j\omega)^{n_0+1}} \times$$
$$\left( \underline{b}_{i',j',0} + \overline{b}_{i',j',0} - \left( \underline{b}_{i',j',N-1} + \overline{b}_{i',j',N-1} \right) \exp(-j\overline{\tau}\omega) + \right. \quad (26)$$
$$\left. \sum_{k=1}^{N-1} \left( \underline{b}_{i',j',k} - \underline{b}_{i',j',k-1} + \overline{b}_{i',j',k} - \overline{b}_{i',j',k-1} \right) \exp(-jkh\omega) \right)$$

Now define matrix $D_k = \left[ d_{i',j',k} \right]_{n\times n}$ with the elements:

$$d_{i',j',k} = \begin{cases} \underline{b}_{i',j',0} + \overline{b}_{i',j',0}, & k=0 \\ \underline{b}_{i',j',k} - \underline{b}_{i',j',k-1} + \overline{b}_{i',j',k} - \overline{b}_{i',j',k-1}, & 1 \le k \le N-1 \\ -\underline{b}_{i',j',N-1} - \overline{b}_{i',j',N-1}, & k=N \end{cases} \quad (27)$$

According to (26), the centers of the eigenvalue inclusion rectangles introduced in Lemma 3 are as follows:

$$Q_C(\omega) = n_0! \sum_{k=0}^{N} tr(D_k) \exp(-jkh\omega) / 2n(j\omega)^{n_0+1} \quad (28)$$

whose real and imaginary parts are given by

$$\text{Re}\{Q_C(\omega)\} = \frac{n_0!(-1)^{1+n_0/2}}{2n\omega^{n_0+1}} \times$$
$$\sum_{k=0}^{N} tr(D_k) \sin(kh\omega)$$
$$\text{Im}\{Q_C(\omega)\} = \frac{n_0!(-1)^{1+n_0/2}}{2n\omega^{n_0+1}} \times \quad (29)$$
$$\sum_{k=0}^{N} tr(D_k) \cos(kh\omega)$$

if $n_0$ is even, and by

$$\text{Re}\{Q_C(\omega)\} = \frac{n_0!(-1)^{(n_0+1)/2}}{2n\omega^{n_0+1}} \times$$
$$\sum_{k=0}^{N} tr(D_k) \cos(kh\omega)$$
$$\text{Im}\{Q_C(\omega)\} = \frac{n_0!(-1)^{(n_0-1)/2}}{2n\omega^{n_0+1}} \times \quad (30)$$
$$\sum_{k=0}^{N} tr(D_k) \sin(kh\omega)$$

otherwise. As using Lemma 1 together with Lemma 3 suggests, encirclements of $Q_C(\omega)$ should be examined in order to decide on the stability status of the system. However, this is only valid provided that the point $+1$ on the complex plane is not trapped within the band $Q$ introduced in Lemma 3. This condition is met if either the inequality

$$\left| \text{Re}\{Q_C(\omega)\} - 1 \right| > \delta_R(\omega)/2 \quad (31)$$

or

$$\left| \text{Im}\{Q_C(\omega)\} \right| > \delta_I(\omega)/2 \quad (32)$$

is satisfied for each $\omega \in \mathbb{R}$. The necessary condition in this respect is $\rho\{T\} < 2$. Fortunately as it is indicated in the next lemma, by using the Biernacki, Pidek, and Ryll-Nardzewski (BPR) inequality it is sufficient to examine condition (31) only in a small frequency range due to the vanishing nature of the functions that are present there.

**Lemma 4**

Let $\rho\{T\} < 2$ and define:

$$\bar{D} = \left[\max_k \{d_{i',j',k}\}\right]_{n \times n}$$

$$\underline{D} = \left[\min_k \{d_{i',j',k}\}\right]_{n \times n}$$

$$\tilde{D} = \bar{D} - \underline{D}$$

$$\tilde{d} = \left(\max_k \{tr(D_k)\} - \min_k \{tr(D_k)\}\right) / 2n$$

where $d_{i',j',k}$ is given by (27). Condition (31) is met for all $\omega \in \mathbb{R}$ if it is satisfied in the range $0 \leq \omega \leq \bar{\omega}$ where $\bar{\omega}$ is given by:

$$\bar{\omega} = \sqrt[n_0+1]{\frac{n_0!(1+N)}{4 - 2\rho\{T\}}\left(2\tilde{d} + \sqrt{\frac{2n-1}{n} tr(\tilde{D}^2 + \tilde{D}^T \tilde{D})}\right)} \quad (33)$$

Proof:

We will find $\omega^* > 0$ such that inequality (31) is met for all frequencies in the range $\omega \in (\omega^*, +\infty)$. Firstly note that since $\sum_{k=0}^{N} tr(D_k) = 0$, the BPR inequality (Cerone & Dragomir (2010), p.9) reveals that the modulus of the real part of $Q_C(\omega)$ satisfies:

$$\left|\text{Re}\{Q_C(\omega)\}\right| \leq 2\tilde{d} C(1+N)(1+N) n_0! / \omega^{n_0+1} \quad (34)$$

in which

$$C(1+N) = \frac{1}{N+1}\left\lfloor\frac{N+1}{2}\right\rfloor\left(1 - \frac{1}{N+1}\left\lfloor\frac{N+1}{2}\right\rfloor\right) \quad (35)$$

By using the reverse triangle inequality and inequality (34) to obtain a lower bound for the left side of (31), it is concluded that inequality (31) is satisfied if:

$$\omega^{n_0+1} > 2n_0! C(1+N)(1+N)\tilde{d} + \omega^{n_0+1}\delta_R(\omega)/2 \quad (36)$$

Next, an upper bound for $\delta_R(\omega)$ is found in a similar way. By expanding the expression of $\delta_R(\omega)$ given by (18) in terms of $\hat{m}_{i',j'}(j\omega)$ given by (26) and using BPR inequality (Cerone & Dragomir (2010), p.9) to bound the moduli of both the real and imaginary parts of $\hat{m}_{i',j'}(j\omega)$, it is eventually deduced that:

$$\delta_R(\omega) \leq \rho\{T\} + ((2n-1)/n)^{1/2} \times$$
$$\left(\text{tr}\left((\hat{M}_R(\omega) + \hat{M}_R^T(\omega))^2 - (\hat{M}_I(\omega) - \hat{M}_I^T(\omega))^2\right)\right)^{1/2} \leq$$
$$\rho\{T\} + ((2n-1)/n)^{1/2} \times \quad (37)$$
$$\left(4\left(n_0! C(1+N)(1+N)/\omega^{n_0+1}\right)^2 \times\right.$$
$$\left. \left(tr(\tilde{D}^2) + tr(\tilde{D}^T \tilde{D})\right)\right)^{1/2} = \rho\{T\} +$$
$$\frac{2n_0!}{\omega^{n_0+1}} C(1+N)(1+N)\sqrt{(2n-1) tr(\tilde{D}^2 + \tilde{D}^T \tilde{D})/n}$$

Finally, using (37) in (36) and noting the inequality $C(1+N) \leq 1/4$ (Cerone & Dragomir (2010), p.9) gives:

$$\omega^{n_0+1} > \frac{n_0!(1+N)}{4 - 2\rho\{T\}}\left(2\tilde{d} + \sqrt{(2n-1) tr(\tilde{D}^2 + \tilde{D}^T \tilde{D})/n}\right) \quad (38)$$

which proves this lemma. ∎

Lemma 4 asserts that if condition (31) or (32) holds for every $\omega \in [0, \bar{\omega}]$, then encirclements of the band $Q$ may be counted without trapping the point $+1$ inside, exposing the exact number of unstable characteristic roots of system (1). Conditions (31)-(32) determine whether the algorithm presented in section 4 is capable of handling the problem. These conditions depend on the system uncertainty and kernel eigenvalues dispersion. In the next theorem, a quick sufficient condition of instability is provided that shall be checked before further investigating the encirclements.

**Theorem 2**

Let (31) or (32) hold in the range $\omega \in [0, \bar{\omega}]$ where $\bar{\omega}$ is given by (33). System (1) is unstable if $tr(\hat{M}(0)) > n$.

Proof:

Following the arguments made in the proof of Theorem 1, it can be shown that every eigenvalue of $M(j\omega)$ makes an equal odd number of encirclements, provided that the assumptions in the statement of this theorem hold. Thus according to Lemma 1, it is realized that system (1) is unstable.

∎

Note that the case $tr(\hat{M}(0)) = n$ is impossible under the condition (31) or (32). Therefore based on Theorem 2, if condition (31) or (32) is met, stability of system (1) needs to be studied further only if:

$$tr(\hat{M}(0)) < n \quad (39)$$

Now the encirclement detection problem is addressed and it is shown that thanks to the spline nature of the kernel bounds, counting the encirclements of the band on the complex plane can be reduced to solving a trigonometric polynomial equation or equivalently an ordinary polynomial equation in Chebyshev form in a finite range, for which various fast methods exist. The general procedure taken here is to find the roots of the imaginary part of $Q_C(\omega)$ with odd multiplicities where the curve $\{Q_C(\omega) \mid \omega \in \mathbb{R}^+\}$ crosses the real axis. Evaluation of the real part of $Q_C(\omega)$ at these roots is a key step in detecting the encirclements and therefore, the stability status of system (1). Therefore in this section, evaluation of $Q_C(\omega)$ in the whole frequency range is reduced to a few key frequency points. In order to clarify and for sake of convenience define:

$$x = h\omega,$$

$$f_k = \begin{cases} tr(D_k) n_0! h^{n_0+1} (-1)^{1+n_0/2} / 2n, & n_0 : \text{even} \\ tr(D_k) n_0! h^{n_0+1} (-1)^{(n_0-1)/2} / 2n, & n_0 : \text{odd} \end{cases} \quad (40)$$

where matrix $D_k$ elements is given in (27). After rewriting the imaginary parts of (29) and (30) in terms of the parameters introduced in (40), it is revealed that in order to find the roots of $\text{Im}\{Q_C(\omega)\} = 0$ we need to consider the equation:

$$y(x) = \sum_{k=1}^{N} f_k \sin(kx) = 0 \quad (41)$$

if $n_0$ is odd, and the equation

$$y(x) = \sum_{k=0}^{N} f_k \cos(kx) = 0 \quad (42)$$

otherwise, if $x \neq 0$. The point $x = 0$ is of course always considered a root since $\text{Im}\{Q_C(0)\} = 0$. Note that if $x = x^*$ is a root of equation $y(x) = 0$ where $x^* \in (0, \pi]$, then $x = x^* + 2k'\pi$ and $x = 2k'\pi - x^*$ are also roots of $y(x) = 0$ where $k' \in \mathbb{N}$. This indicates that the roots of $y(x) = 0$ in the range $(0, \pi]$ need to be found only, which can be calculated immediately by using Fourier companion matrices as stated in the following lemma.

**Lemma 5**

Define:

$$F_{2N \times 2N} = \begin{bmatrix} 0_{(2N-1) \times 1} & I_{(2N-1) \times (2N-1)} \\ -\frac{f_N}{f_N} & \cdots & -\frac{f_1}{f_N} & -\frac{2f_0}{f_N} & -\frac{f_1}{f_N} & \cdots & -\frac{f_{N-1}}{f_N} \end{bmatrix} \quad (43)$$

if $n_0$ is even, and

$$F_{2N \times 2N} = \begin{bmatrix} 0_{(2N-1) \times 1} & I_{(2N-1) \times (2N-1)} \\ \frac{f_N}{f_N} & \cdots & \frac{f_1}{f_N} & 0 & -\frac{f_1}{f_N} & \cdots & -\frac{f_{N-1}}{f_N} \end{bmatrix} \quad (44)$$

otherwise. Let $\lambda_i$ ($i = 1, 2, \cdots, \alpha$) denote the eigenvalues of $F$ on the upper semi-circular part of the unitary circle including its end points $\pm 1$ with odd algebraic multiplicities. Define $x_i = \angle \lambda_i$. Then $x = x_i$ ($i = 1, 2, \cdots, \alpha$) satisfies $y(x) = 0$ in the range $[0, \pi]$.

Proof:

Matrix $F$ defined by (43) or (44) is in fact a Fourier companion matrix in CCM form. The CCM form of companion matrices is used here due to its simplicity and numerical efficiency among other root-finding methods for deriving all roots of trigonometric polynomial equations (Boyd (2014)). For more details one may consult Boyd (2014). ∎

The next step would be to evaluate the real parts $\text{Re}\{Q_C(x/h)\}$ at the roots $x = x_i$. Then the real axis crossovers by the centers of $Q(\omega)$ can be detected, which in turn would expose the encirclements in a relatively cheap procedure. Therefore having calculated the sequence

$$X_i = \begin{cases} 0, & i = 0 \\ tr(\hat{M}(jx_i/h))/n, & i = 1, \cdots, \alpha \end{cases} \quad (45)$$

robust stability problem of system (1) can be tackled by using the following theorem.

**Theorem 3**

Assume that (39) is met and condition (31) or (32) holds for every $\omega \in [0, \bar{\omega}]$ where $\bar{\omega}$ is given by (33). Define $J = 2\alpha \left\lfloor \left( \sqrt[n_0+1]{\max_{1 \leq i \leq \alpha} \{|X_i|\}} + 1 \right)/2 \right\rfloor + \alpha$ and the sequences:

$$
\begin{aligned}
g_i &= \left(1 - (-1)^{\lfloor (i-1)/\alpha \rfloor}\right)(\alpha+1)/2 + \\
&\quad (-1)^{\lfloor (i-1)/\alpha \rfloor} \left(i - \lfloor (i-1)/\alpha \rfloor \alpha\right) \\
x_i &= (-1)^{\lfloor (i-1)/\alpha \rfloor} x_{g_i} + 2\pi \lceil \lfloor (i-1)/\alpha \rfloor /2 \rceil \\
X_i &= \left((-1)^{\lfloor (i-1)/\alpha \rfloor} x_{g_i}/x_i\right)^{n_0+1} X_{g_i} \\
\upsilon_i &= (X_i - 1)(X_{i+1} - 1) \\
\mu_i &= X_{i+1} - X_i
\end{aligned}
\qquad (46)
$$

where $1 \leq i \leq J$ with the initial samples of $X_i$ and $x_i$ as given by (45) and in Lemma 5 respectively. Characteristic equation (6) has exactly $\zeta$ number of unstable roots in the open right half plane, where:

$$
\zeta = n \left| \sum_{0 \leq i \leq J,\; \upsilon_i < 0} (-1)^i \operatorname{sgn}(\mu_i) \right| \qquad (47)
$$

Proof:

After deriving the roots $x = x_i$ ($i = 1, 2, \cdots, \alpha$) of $y(x) = 0$ in the range $[0, \pi]$ by using Lemma 5, due to the periodicity and symmetry of $y(x)$, all the roots on the non-negative real axis are constructed as:

$$ x_i = \operatorname{sgn}(\hat{k})\left(x_{g_i} + 2\hat{k}\pi\right) \qquad (48) $$

where $\hat{k} \in \mathbb{Z}$, $i \in \mathbb{N}$, $1 \leq g_i \leq \alpha$. Relation (48) extends the sequence $x_i$ for all $i \in \mathbb{N}$. Thanks to the spline nature of the bounding kernels, $X_i = Q_C(x/h)\big|_{x = x_i}$ may be recursively calculated using the following relation:

$$ X_i = \left(\operatorname{sgn}(\hat{k}) x_{g_i}/x_i\right)^{n_0+1} X_{g_i} \qquad (49) $$

where $\hat{k} \in \mathbb{Z}$, $i \in \mathbb{N}$, $1 \leq g_i \leq \alpha$, due to the symmetry and periodicity in the trigonometric polynomials associated with the real parts in (29) and (30). Defining

$$
\begin{aligned}
\hat{k} &= (-1)^{\lfloor (i-1)/\alpha \rfloor} \lceil \lfloor (i-1)/\alpha \rfloor /2 \rceil \\
g_i &= (1 - \operatorname{sgn}(\hat{k}))(\alpha+1)/2 + \\
&\quad \operatorname{sgn}(\hat{k})\left(i - \lfloor (i-1)/\alpha \rfloor \alpha\right)
\end{aligned}
\qquad (50)
$$

in (48) sorts the sequence $x_i$ by ascending size, which yields in definition (46). Now let us separate the complex plane $\mathbb{C} \setminus \{1\}$ into the following regions:

1. $\phi_+ = \{z \in \mathbb{C} \mid \operatorname{Im}(z) > 0\}$

2. $\phi_0 = \phi_{1-} \cup \phi_{1+}$, where

$$
\begin{cases}
\phi_{1-} = \{z \in \mathbb{C} \mid \operatorname{Im}(z) = 0,\; \operatorname{Re}(z) < 1\} \\
\phi_{1+} = \{z \in \mathbb{C} \mid \operatorname{Im}(z) = 0,\; \operatorname{Re}(z) > 1\}
\end{cases}
$$

3. $\phi_- = \{z \in \mathbb{C} \mid \operatorname{Im}(z) < 0\}$

through which the centers of the rectangles $Q(\omega)$ may travel. For positive frequencies $\omega > 0$ (or equivalently $x > 0$), it is only when $x = x_i$ that $Q_C(x/h)$ switches between the regions $\phi_+ \cup \phi_0$ and $\phi_- \cup \phi_0$ in which $x_i$ is defined by (46). At $x = 0$, one has $Q_C(x/h) \in \phi_{1-}$. Define $s_0$ as a two-valued variable in a way that $s_0 = 1$ holds if $Q_C(x/h) \in \phi_+ \cup \phi_0$, and $s_0 = -1$ holds if $Q_C(x/h) \in \phi_- \cup \phi_0$ in the range $x \in (0, x_1)$ before the first switch occurs. Hence in the range $x \in (x_i, x_{i+1})$, one has $Q_C(x/h) \in \phi_- \cup \phi_0$ ($Q_C(x/h) \in \phi_+ \cup \phi_0$) for $i$ being an odd (even) number if $s_0 = 1$, and the opposite situation holds if $s_0 = -1$. Based on this argument, it is deduced that $\operatorname{sgn}(\operatorname{Im}\{Q_C(x/h)\}) = s_0 (-1)^i$ holds for almost all $x \in (x_i, x_{i+1})$. A half-encirclement of the point $+1$ is occurred in the range $x \in (x_i, x_{i+1})$ when $1 \in (\min\{X_i, X_{i+1}\}, \max\{X_i, X_{i+1}\})$ or equivalently when $\upsilon_i < 0$ in which $\upsilon_i$ is defined by (46). The direction of this half-encirclement can be easily determined by $s_0 (-1)^i$ together with $\operatorname{sgn}(\mu_i)$ where $\mu_i$ is defined by (46). Therefore, all encirclements of the point $+1$ occurring in the positive frequencies range is eventually given by:

$$\frac{1}{2} \sum_{1 \leq i,\ \upsilon_i < 0} s_0 (-1)^i \operatorname{sgn}(\mu_i) \tag{51}$$

Therefore the total number of encirclements in the range $\omega \in (-\infty, +\infty)$ which is always non-negative according to Lemma 1, is given by:

$$\zeta = n \left| \sum_{1 \leq i,\ \upsilon_i < 0} (-1)^i \operatorname{sgn}(\mu_i) \right| \tag{52}$$

Fortunately due to the attenuation term $1/x^{n_0+1}$ in (29) and (30), the summation above always has a finite number of terms, since for some $J \in \mathbb{N}$ there holds $\upsilon_i > 0$ for $i > J > \alpha$. In order to obtain $J$, one may solve the inequality $|X_i| < 1$ for $i$. Thus from (49) and (48) one may write:

$$x_{g_i} \left( \sqrt[n_0+1]{|X_{g_i}|} - \operatorname{sgn}(\hat{k}) \right) / 2\pi < |\hat{k}| \tag{53}$$

assuming $|\hat{k}| > 0$ as $J > \alpha$. Inequality (53) is fulfilled provided that:

$$x_{g_i} \left( \sqrt[n_0+1]{\max_{1 \leq i \leq \alpha} \{|X_i|\}} + 1 \right) / 2\pi < |\hat{k}| \tag{54}$$

holds true. Define $\bar{k} = \left( \sqrt[n_0+1]{\max_{1 \leq i \leq \alpha} \{|X_i|\}} + 1 \right) / 2$. Since there holds $0 \leq x_{g_i} \leq \pi$, inequality (54) is satisfied if $\bar{k} < |\hat{k}|$. In order to find a corresponding bound for $i$, definition (50) can be used to give $\bar{k} < \lceil \lfloor (i-1)/\alpha \rfloor / 2 \rceil$. Proceeding with this inequality results in $i \geq 2\alpha \lfloor \bar{k} \rfloor + 1$ if $\lfloor (i-1)/\alpha \rfloor$ is odd, and in $i \geq 2\alpha \lfloor \bar{k} \rfloor + \alpha + 1$ otherwise. Therefore $\upsilon_i > 0$ is met for $i > J$, provided that $J$ is defined as in the statement of this theorem. ∎

Based on Theorem 3, system (1) is obviously robust stable if $\zeta = 0$.

## 4. Algorithm

For a convenient use, an algorithm for robust stability analysis of system (1) is provided in this section based on the methodology presented in this paper. This algorithm facilitates the implementation of obtained results in computer programs and it is designed by mainly manipulating theorem 3 to make it even cheaper. If (39) and condition (31) or (32) hold, the following algorithm successfully gives the exact number of unstable characteristic roots of the uncertain system (1).

**Algorithm:**

1. Calculate the spectral radius of $T$ as given by (18). If $\rho\{T\} \geq 2$, then the stability analysis is inconclusive for an unbearable uncertainty. Otherwise, proceed to next step.

2. Calculate $\bar{\omega}$ as given by (33). If at least one of the conditions (31) or (32) holds for every $\omega \in [0, \bar{\omega}]$, proceed to next step. Stability analysis is inconclusive otherwise.

3. If $tr(\hat{M}(0)) > n$, then system (1) is unstable having an odd multiple of $n$ unstable roots. Otherwise proceed to next step.

4. Compute coefficients $f_k$ by using (40) and form matrix $F$ defined in Lemma 5.

5. Compute $x_i$ ($i = 1, 2, \cdots, \alpha$) from Lemma 5 using matrix $F$. Set $X_0 = 0$ and compute $X_i$ ($i = 1, 2, \cdots, \alpha$) using (45). If $\alpha = 0$ or $\max_{1 \leq i \leq \alpha} \{|X_i|\} < 1$ holds, then system (1) is robust stable. Otherwise, proceed to next step.

6. Define $Z_i = X_i$ for $i = 0, 1, 2, \cdots, \alpha$ and let $\beta = 1$, $J' = 1$, $\gamma(-1) = 0$, $\gamma(+1) = 0$.

7. Count the number of jumps present in the sequence $Z_i$ from $i = 0$ to $i = \alpha$. This is done in a way that if there is a descent from $Z_i$ to $Z_{i+1}$ crossing the threshold value of $1$, there is a negative (positive) jump if $i + (1 - \beta)\alpha / 2$ is even (odd). Likewise, if an ascent from $Z_i$ to $Z_{i+1}$ crossing the threshold value of $1$ occurs, then there is a positive (negative) jump if $i + (1 - \beta)\alpha / 2$ is even (odd).

8. Set $Z_0 \leftarrow Z_\alpha$.

9. If $\beta = -1$, flip the sequence $Z_i$, ($i = 1, 2, \cdots, \alpha$).

10. If $|Z_i| > 1$, update

$$Z_i \leftarrow X_i \left( \frac{x_i}{x_i - 2\pi\beta J'} \right)^{n_0+1}$$

for $1 \leq i \leq \alpha$.

11. If $\beta = +1$, flip the sequence $Z_i$, ($i = 1, 2, \cdots, \alpha$).

12. Set $\beta \leftarrow (-\beta)$.

13. Set $J' \leftarrow J' + (1+\beta)/2$.

14. If there are no samples greater than 1 left in the sequence $Z_i$, ($i = 0, 2, \cdots, \alpha$) set $\gamma(\beta) = 1$.

15. If $\gamma(-1) = 1$ and $\gamma(+1) = 1$, add the total number of recorded jumps together, calculate its absolute value and multiply the result by $n$. This gives the exact number of unstable roots in the open right half plane. Otherwise, return to step 7.

## 5. Numerical evaluations

In this section, three numerical examples are presented to examine the efficiency of the results obtained in this paper. In the first two examples, conservatism of the proposed method is compared with some previous results available in the literature. Since robust stability is concluded almost immediately in the early steps of the algorithm in the first two examples, a third example is also provided to fully assess the algorithm capabilities through all its steps.

**Example 1**

Consider system (1) with $\bar{\tau} = 0.5$ and the exponential kernel matrix $A(\tau) = \exp(-A'\tau)$ where $A' = \begin{bmatrix} 0 & 1 \\ 0 & 0 \end{bmatrix}$. This system was shown to be stable by Mondié & Melchor-Aguilar (2012). In this example, it is also assumed that the nominal kernel $\exp(-A'\tau) = \begin{bmatrix} 1 & -\tau \\ 0 & 1 \end{bmatrix}$ is subject to some perturbations. Robustness of this stable system is then assessed by using the algorithm of section 4, assuming that the kernel satisfies (2) where:

$$\bar{A}(\tau) = \begin{bmatrix} 1.2 & \bar{a}_{1,2}(\tau) \\ 0.2 & 1.2 \end{bmatrix}, \underline{A}(\tau) = \begin{bmatrix} 0.8 & \underline{a}_{1,2}(\tau) \\ -0.2 & 0.8 \end{bmatrix}$$

and

$$\bar{a}_{1,2}(\tau) = -\sum_{k=0}^{4} k \left( H(\tau - kh) - H(\tau - (k+1)h) \right)$$

$$\underline{a}_{1,2}(\tau) = -\sum_{k=0}^{4} (k+1) \left( H(\tau - kh) - H(\tau - (k+1)h) \right)$$

Since $\rho\{T\} = 0.65$, we proceed to calculate $\bar{\omega} = 30.87$ and confirm that both (39) and (31) are satisfied. Proceeding with the algorithm proves the system robust stable with no update session (step 10) required.

**Example 2**

Consider system (1) with $\bar{\tau} = 2$ and the kernel $A(\tau) = -c_1 \tau + c_2$ which describes internal dynamics of the controllers used for assigning finite spectrums to the closed loop version of the input-delay systems (Morales-Sánchez & Melchor-Aguilar (2013)). Let the nominal gains be $\hat{c}_1 = -0.0005$, $\hat{c}_2 = -0.0267$ and assume that the gains vector $[c_1 \; c_2]$ is unknown within a certain disk with its center located at the nominal gains $[\hat{c}_1 \; \hat{c}_2]$ and radius $r_c = 0.1439$. This system was proved robust stable in Morales-Sánchez & Melchor-Aguilar (2013). In this example however, we use greater upper bounds and smaller lower bounds of the perturbed kernel and allow kernel to be any function of $\tau$ (not just linear) and yet prove the system robust stable by using the algorithm of section 4. Then it is shown that robust stability of the proposed system is also concluded even after the perturbation radius $r_c$ has also been increased.

It can be shown that the perturbed kernel $A(\tau)$ sweeps the surface between the two curves:

$$\bar{B}(\tau) = -\hat{c}_1 \tau + \hat{c}_2 + r_c \sqrt{\tau^2 + 1}$$
$$\underline{B}(\tau) = -\hat{c}_1 \tau + \hat{c}_2 - r_c \sqrt{\tau^2 + 1}$$

Due to the monotonicity nature of $\bar{B}(\tau)$ and $\underline{B}(\tau)$ in the range $\tau \in [0.0035, 2]$, the elements of kernel bounds $\underline{A}(\tau)$ and $\bar{A}(\tau)$ can be simply defined as in (4) by choosing $n_0 = 0$, $h = 2/3$, $\bar{b}_{i',j',k} = \bar{B}(kh + h)$ and $\underline{b}_{i',j',k} = \underline{B}(kh + h)$. In this setting, $\rho\{T\} = 1.96 < 2$, condition (39) is met and the first condition in (31) is satisfied for $0 \le \omega \le \bar{\omega}$ where $\bar{\omega} = 12.2$. Therefore, stability status of the system can be certainly determined through the algorithm of section 4. It is observed that $\alpha = 0$ and therefore the proposed system is proved robust stable in step 5 which obviates the need to further proceed with the algorithm altogether. Now three numerical experiments are performed by having $h = 0.1$ fixed and choosing different values of $\bar{\tau} \in \{2, 5, 10\}$. In each case the corresponding perturbation radius $r_c'$ for which the system is proved robust stable by using the algorithm of this paper is recorded in table 1. Results of similar experiments can be found in Morales-Sánchez & Melchor-Aguilar (2013) making it possible to compare the results. It is worth mentioning that the proposed system is proved robust stable in the step 5 of the algorithm for $\bar{\tau}=2$ and $\bar{\tau}=5$, and with no sample updates (step 10) needed for $\bar{\tau}=10$.

| $\bar{\tau}$ | 2 | 5 | 10 |
|---|---|---|---|
| $r_c$ | 0.1439 | 0.0219 | 0.00441 |
| $r_c'$ | 0.16 | 0.035 | 0.0095 |
| $100\times(r_c'-r_c)/r_c$ | +11.19% | +59.82% | +115.42% |

**Table 1:** The perturbation radius $r_c'$ for which the system is proved robust stable by using the algorithm of this paper versus the perturbation radius $r_c$ for which the system is proved robust stable in Morales-Sánchez & Melchor-Aguilar (2013).

As it is apparent in Table 1, the proposed algorithm in section 4 has lower conservatism in all experiments.

**Example 3**

Consider system (1) with $n=1$, $n_0=1$, $h=0.5$, $\bar{\tau}=1$, $\bar{b}_{1,1,0}=\underline{b}_{1,1,0}=-30$ and $\bar{b}_{1,1,1}=\underline{b}_{1,1,1}=30$. Since this system contains no uncertainty, one has $\rho\{T\}=0$. Using (33), $\bar{\omega}$ is calculated as $\bar{\omega}=18.05$. It can be shown that (31) holds for $0\leq\omega\leq\bar{\omega}$. Also since $tr(\hat{M}(0))=-15/2$, by proceeding with the algorithm of section 4 the exact number of unstable characteristic roots of (1) can be obtained. Proceeding from step 4 to step 5 gives $x_1=\pi$, $x_2=0$, $x_3=2.8\times10^{-6}$ which implies $\alpha=3$. Therefore we need to calculate $X_1=3.0396$, $X_2=-7.5$ and $X_3=-7.49999999997$. Following the algorithm in step 7 one ascent and one descent is recorded which are interpreted as two positive jumps. In the update session (step 10), three samples are updated. In step 14, $\gamma(\pm1)$ doesn't change and we return to step 7 with the updated sequence $Z_0=-7.5$, $Z_1=0$, $Z_2=0$ and $Z_3=3.0396$. This time, one ascent as a positive jump is recorded and in the update session one sample is updated. Again, $\gamma(\pm1)$ isn't changed in this round and we return to step 7. In the next round there is recorded a descent as a negative jump in step 7 and no more samples are updated in step 10. In step 14, it is set $\gamma(-1)=1$. Returning to step 7, no more jumps or updates are occurred and it is set $\gamma(+1)=1$ in step 14. Therefore, the algorithm stops with the conclusion that the proposed system is stable as the total number of jumps added together is zero.

## 6. Conclusion

In this paper a fully algebraic algorithm was presented based on the characteristic equations and the argument principle for robust stability analysis of integral delay systems with uncertain kernels. The intrinsic complexity of characteristic equations of integral delay systems as well as their complicated Nyquist diagrams with a large number of crossings and encirclements are among the main reasons for which frequency domain-based methods for these systems are not so popular. Therefore, the provided algorithm should be a contribution to overcome this shortcoming by reducing the computation demand. This algorithm tries to use the advantage of low conservatism involved in the frequency domain analysis, while remaining relatively simple and cheap with an easy implementation. The presented method shows the best results for scalar systems, diagonally dominant systems and systems with close eigenvalues. As it was shown, the presented algorithm is less conservative than the existing methods compared in the considered examples. Moreover despite the available methods as Ortiz et al. (2018), it can handle kernels of general types, not being limited to just exponential and constant kernels.

**References**

Berry, R. A., & Gallager, R. G. (2002). Communication over fading channels with delay constraints. *IEEE Transactions on Information Theory*, *48*(5), 1135-1149.

Boyd, J. P. (2014). *Solving Transcendental Equations: The Chebyshev Polynomial Proxy and Other Numerical Rootfinders, Perturbation Series, and Oracles*. Society for Industrial and Applied Mathematics.

Cerone, P., & Dragomir, S. S. (2010). *Mathematical inequalities: a perspective*. CRC Press.

Chen, D., & Seborg, D. E. (2002). Robust Nyquist array analysis based on uncertainty descriptions from system identification. *Automatica,* 38(3), 467-475.

Hladík, M. (2013). Bounds on eigenvalues of real and complex interval matrices. *Applied Mathematics and Computation*, *219*(10), 5584-5591.

Juang, Y. T., & Shao, C. S. (1989). Stability analysis of dynamic interval systems. *International Journal of Control*, *49*(4), 1401-1408.

Li, Z. Y., Zheng, C., & Wang, Y. (2016). Exponential stability analysis of integral delay systems with multiple exponential kernels. *Journal of the Franklin Institute*, *353*(7), 1639-1653.

Li, Z. Y., Zhou, B., & Lin, Z. (2013). On exponential stability of integral delay systems. *Automatica*, *49*(11), 3368-3376.

Maciejowski, J. M. (1989). Multivariable feedback design. Electronic systems engineering series. *Wokingham, England: Addison-Wesley,* 6, 85-90.

Mehrkanoon, S., Shardt, Y. A., Suykens, J. A., & Ding, S. X. (2016). Estimating the unknown time delay in chemical processes. *Engineering Applications of Artificial Intelligence*, *55*, 219-230.

Melchor-Aguilar D., Morales-Sánchez A. (2016) Robust Stability of Integral Delay Systems with Exponential Kernels. In: Witrant E., Fridman E., Sename O., Dugard L. (eds) Recent Results on Time-Delay Systems. Advances in Delays and Dynamics, vol 5. Springer, Cham.

Melchor-Aguilar, D. (2016). New results on robust exponential stability of integral delay systems. *International Journal of Systems Science*, *47*(8), 1905-1916.

Melchor-Aguilar, D., Kharitonov, V., & Lozano, R. (2008, December). Stability and robust stability of integral delay systems. In *Decision and Control, 2008. CDC 2008. 47th IEEE Conference on* (pp. 4640-4645). IEEE.

Melchor-Aguilar, D., Kharitonov, V., & Lozano, R. (2010). Stability conditions for integral delay systems. *International Journal of Robust and Nonlinear Control*, *20*(1), 1-15.

Mondié, S., & Melchor-Aguilar, D. (2012). Exponential stability of integral delay systems with a class of analytic kernels. *IEEE Transactions on Automatic Control*, *57*(2), 484-489.

Morales-Sánchez, A., & Melchor-Aguilar, D. (2013). Robust stability conditions for integral delay systems with exponential kernels. *IFAC Proceedings Volumes*, *46*(3), 132-137.

Ortiz, R., Mondié, S., Del Valle, S., & Egorov, A. V. (2018, December). Construction of Delay Lyapunov Matrix for Integral Delay Systems. In *2018 IEEE Conference on Decision and Cont*

Si, W., Qi, L., Hou, N., & Dong, X. (2019). Finite-time adaptive neural control for uncertain nonlinear time-delay systems with actuator delay and full-state constraints. *International Journal of Systems Science*, *50*(4), 726-738.

Sun, H., Hou, L., Zong, G., & Yu, X. (2019). Adaptive Decentralized Neural Network Tracking Control for Uncertain Interconnected Nonlinear Systems With Input Quantization and Time Delay. *IEEE transactions on neural networks and learning systems*.

Taghavian, H., & Tavazoei, M. S. (2017). Robust Stability Analysis of Distributed-Order Linear Time-Invariant Systems With Uncertain Order Weight Functions and Uncertain Dynamic Matrices. *Journal of Dynamic Systems, Measurement, and Control*, *139*(12), 121010.

Taghavian, H., & Tavazoei, M. S. (2018). Robust stability analysis of uncertain multiorder fractional systems: Young and Jensen inequalities approach. *International Journal of Robust and Nonlinear Control*, *28*(4), 1127-1144.

Vafamand, N., Khooban, M. H., Dragicevic, T., Blaabjerg, F., & Boudjadar, J. (2019). Robust non-fragile fuzzy control of uncertain DC microgrids feeding constant power loads. *IEEE Transactions on Power Electronics*.

Yan, H., Wang, J., Wang, F., Wang, Z., & Zou, S. (2019). Observer-based reliable passive control for uncertain T–S fuzzy systems with time-delay. *International Journal of Systems Science*, *50*(5), 905-918.

Zhou, B., & Li, Z. Y. (2016). Stability analysis of integral delay systems with multiple delays. *IEEE Transactions on Automatic Control*, *61*(1), 188-193.

Zou, L., & Jiang, Y. (2010). Estimation of the eigenvalues and the smallest singular value of matrices. *Linear Algebra and its Applications*, *433*(6), 1203-1211.